\documentclass[doublecol]{epl2}
\revision

\title{Results in Kalb-Ramond field localization and resonances on deformed branes}
\shorttitle{Kalb-Ramond field localization and resonances} %Insert here a short version of the title if it exceeds 70 characters

\author{W. T. Cruz\inst{1,3} \and M. O. Tahim\inst{2} \and C. A. S. Almeida\inst{3}}
\shortauthor{W.T. Cruz \etal}

\institute{
  \inst{1}Instituto Federal de Educa\c{c}\~{a}o, Ci\^{e}ncia e Tecnologia do Cear\'{a} (IFCE), Campus Juazeiro do Norte - 63040-000 Juazeiro do Norte-Cear\'{a}-Brazil\\
\inst{2} Departamento de Ci\^{e}ncias da Natureza, Faculdade de Ci\^{e}ncias, Educa\c{c}\~{a}o e Letras
do Sert\~{a}o Central (FECLESC), Universidade Estadual do Cear\'{a} - 63900-000
Quixad\'{a}-Cear\'{a}-Brazil \\
  \inst{3} Departamento de F\'{i}sica - Universidade Federal do Cear\'{a} - C.P. 6030, 60455-760
Fortaleza-Cear\'{a}-Brazil}


\abstract{
We make an analysis about several aspects of localization
of the Kalb-Ramond gauge field in a specific four
dimensional AdS membrane embedded in a five dimensional space-time.
The membrane is generated from a deformation of the $\lambda \phi^4$
potential and belongs to a new class of defect solutions. In this context we find resonance structures in the analysis of massive modes. The study
of deformed defects is important because they contain internal
structures and these may have implications to the way the background
space-time is constructed and the way its curvature behaves.  The main
objective here is to observe the contributions of the deformation
procedure to the resonances and the well known field localization methods.}

\begin{document}

\maketitle

\section{Introduction}

In a scenario of extra dimensions the observable universe is represented by a four-dimensional membrane embedded in a higher dimensional space-time. The standard model of particles is confined in the membrane while gravitation is free to propagate into the extra dimension. These ideas have appeared as alternatives to solve the gauge hierarchy problem \cite{RS}. Recently a lot of attention has been given to the study of topological defects in the context of warped space-times. The number of extra dimensions guides us in choosing the right type of defect in order to mimic our brane-world. The key idea for construction of the brane-world is to localize in a very natural way the several fields  of our universe (the bosonic ones and fermionic ones). In this way several works have considered five-dimensional universes \cite{gremm,de} where five dimensional gravity is coupled to scalar fields. In this scenario, with a specific choice for the scalar potential, it is obtained thick domain wall as solutions that may be interpreted as non-singular versions of the Randall-Sundrum scenario. Besides gravity, the study of localization of fields with several spins is very important \cite{kehagias}. Also, this type of scenario contributes for discussions about cosmology. In models with 5-dimensional membranes, the mechanism controlling the expansion of the universe have been associated to the thickness of the membrane along the extra dimension \cite{observations}.

As known, the kind of structure of the considered membrane is very
important and will produce implications concerning the methods of
field localization. In the seminal works of Bazeia and collaborators \cite{deformed,aplications} a class
of topological defect solutions was constructed starting from a
specific deformation of the $\phi^4$ potential. These new solutions
may be used to mimic new brane-worlds containing internal structures
\cite{aplications}. Such internal structures have implications in the
density of matter-energy along the extra dimensions \cite{brane} and
this produces a space-time background whose curvature has a
splitting, as we will show, if compared to the usual models. Some
characteristics of such model were considered in phase transitions
in warped geometries \cite{fase}.

Motivated by the references above, our main subject here is to answer the
following question: Are these structures able to localize the tensor gauge field? In a previous work \cite{nosso}, we find resonances by analyzing the massive spectrum of the Kalb-Ramond field on Bloch branes. Now we analyze the behavior of these structures in a more complex type of membrane. This letter is organized as follows: in the second
section we describe how the deformed membrane is constructed and how
the space-time background is obtained; in the third section we study
the localization of the Kalb-Ramond gauge field in the background obtained; the fourth section is important because we introduce the dilaton field in
order to force the localization of the Kalb-Ramond field. Such analysis is made in the fifth section; in the sixth section we analyze the massive modes using a Supersymmetric Quantum Mechanics formalism; the seventh section deals the resonance structures in the massive spectrum and its relation with deformations. At the final we present our
conclusions and perspectives.

\section{Two-kink solutions modeling the brane}

There is great interest in studying scalar fields coupled to
gravity. If we consider a $D=5$ universe, we should embed a kink
solution in this space-time in order to build our membrane. These
kind of solutions are obtained through the $\lambda\phi^4$ or
sine-Gordon potentials. In our case, following the reference
\cite{deformed}, we will obtain a new class of defects starting from
a deformation of the $\lambda\phi^4$ potential. In this way we can
analyze localization of fields of several ranks in a more complete
fashion because the deformed membranes suggests the existence of
internal structures. As we will see, this choice avoids space-time
singularities also, which is only possible by choosing smooth membrane
solutions.

Our model is built with an AdS $D=5$ space-time whose metric is
given by
\begin{equation}
ds^{2}=e^{2A(y)}\eta_{\mu\nu}dx^{\mu}dx^{\nu}+dy^{2}.
\end{equation}
The warp factor is composed by the function $A(y)$, where $y$ is the
extra dimension. The tensor $\eta_{\mu\nu}$ stands for the Minkowski
space-time metric and the indexes $\mu$ and $\nu$ go from $0$ to
$3$.

In order to construct the membrane solution we start with an action
describing the coupling between a real scalar field and gravitation:
\begin{equation}
S=\int d^{5}x
\sqrt{-G}[2M^{3}R-\frac{1}{2}(\partial\phi)^{2}-V(\phi)].
\end{equation}
In the last action, the field $\phi$ represents the stuff from which
the membrane is made, $M$ is the Planck constant in $D=5$ and $R$ is
the  scalar curvature. The equations of motion coming from that
action are:
\begin{equation}\label{mov1}
\frac{1}{2}(\phi^{\prime})^{2}-V(\phi)=24M^{3}(A^{\prime})^{2},
\end{equation}
\begin{equation}\label{mov2}
\frac{1}{2}(\phi^{\prime})^{2}+V(\phi)=-12M^{3}A^{\prime\prime}-24M^{3}(A^{%
\prime})^{2}.
\end{equation}
Note that the prime means derivative with respect to the extra
dimension. Basically, we look for solutions in which $\phi$ tends to
different values when $y\rightarrow\pm\infty$. In a flat space-time
we find kink-like solutions for the above equations by choosing a
double-well potential. Analogously, if we look for bounce-like
solutions in curved space-time, we should regard potentials
containing various minima. In the presence of gravity, we can find
first order equations by the superpotential method if we take the
superpotential $W(\phi)$ in such a way that $\frac{\partial
W}{\partial \phi}=\phi'$. Our potential must be defined by
\begin{equation}
V_p(\phi)=\frac{1}{2}\left(\frac{dW}{d\phi}\right)^2-\frac{8M^3}{3}W^2,
\end{equation}
from where we can conclude that $W=-3A'(y)$. This formalism was
initially introduced to study non-supersymmetric domain walls in
various dimensions \cite{de,sken}.

Following the references \cite{deformed,aplications,adauto,defects-inside}
the superpotential is given by,
\begin{equation}\label{sup}
W_p(\phi)=\frac{p}{2p-1}\phi^{\frac{2p-1}{p}}-\frac{p}{2p+1}\phi^{\frac{2p+1}{p}},
\end{equation}
where $p$ is an odd integer. The choice for $W_p$ can be obtained by
deforming the usual $\phi^4$ model and it is introduced in the study of
deformed membranes \cite{adauto}. This choice will permit us to get
new and well behaved models for $p=1,3,5,...$ (for $p=1$ we get the
usual $\phi^4$ model). For $p=3,5,7,...,$ the potential $V_p$ has
one minimum at $\phi=0$ and two at $\pm 1$. A new class of solutions
called two-kink solutions initially presented in Ref.\cite{aplications} can be
obtained from the choice of the superpotential $W_p$. For this we
solve $\frac{\partial W}{\partial \phi}=\phi'$ to find
\begin{equation}\label{twokink}
\phi_p(y)=tanh^p(\frac{y}{p}).
\end{equation}
Starting from the first order equation $W_p=-3A_p'(y)$, we can find
the solution for the function $A_p(y)$ \cite{adauto},
\begin{eqnarray}\label{a}
A_p(y)=-\frac{1}{6}\frac{p}{2p+1}\tanh^{2p}\left(\frac{y}{p}\right)\\\nonumber-
\frac{1}{3}\left(\frac{p^2}{2p-1}-\frac{p^2}{2p+1}\right)\\\nonumber
\biggl{\{}\ln\biggl[\cosh\left(\frac{y}{p}\right)\biggr]-
\sum_{n=1}^{p-1}\frac1{2n}\tanh^{2n}\left(\frac{y}{p}\right)\biggr{\}}
\end{eqnarray}
The function $A(y)$ determines the behavior of the warp factor. The
characteristics of localization for several fields and the
construction of effective actions in $D=4$ will depend on part of
the contribution of the warp factor. Note that the exponential
factor constructed with this function is localized around the
membrane and for large $y$ it approximates the Randall-Sundrum
solution \cite{RS}. The solution found here reproduces the
Randall-Sundrum model in an specific limit. The space-time now has
no singularity because we get a smooth warp factor (because of this,
the model is more realistic) \cite{kehagias}. In fact this can be
seen by calculating the curvature invariants for this geometry. For
example, we obtain

% \FIGURE{\centerline{
%\epsfig{figure=curvp.eps,width=15.2cm,height=4.6cm}}
%\caption[caption]{Plots of the solution of the curvature invariant
%$R(y)$$p=1$ (left) and for $p=3$ (dashed line), $p=5$ (doted line)
%and $p=7$ (solid line) (right) .}\label{curvp}}

%\begin{figure}{\centerline{
%\epsfig{figure=curvp.eps,width=15.2cm,height=4.6cm}}
%\caption{Plots of the solution of the curvature invariant
%$R(y)$$p=1$ (left) and for $p=3$ (dashed line), $p=5$ (doted line)
%and $p=7$ (solid line) (right) .}\label{curvp}}
%\end{figure}

\begin{figure}
\onefigure[scale=0.80]{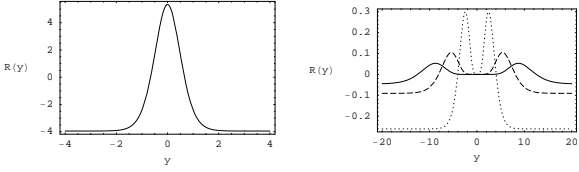}
\caption{Plots of the solution of the curvature invariant
$R(y)$ $p=1$ (left) and for $p=3$ (dashed line), $p=5$ (doted line)
and $p=7$ (solid line) (right).}
\label{curvp}
\end{figure}

\begin{equation}
R=-[8A_p''+20(A_p')^{2}],
\end{equation}
Note that the Ricci scalar is finite, which we can observe through
the Fig.(\ref{curvp}). We can see also an important characteristic
of the deformed structure in comparison with the usual thick
membrane models generated by simple kinks. Observing again the
Fig.(\ref{curvp}), for $p=1$, we have the usual curvature scalar
for the non-deformed model, the usual one. In this case, the
curvature has maximum at $y=0$ and goes to negative values when
$y\rightarrow\infty$. However, regarding the deformed model, taking
$p=3,5$ in $R_p(y)$, we obtain a splitting with the appearance of a
region of zero curvature between two maxima.

\section{The Kalb-Ramond field}

In this section we analyze the behavior of the Kalb-Ramond field in the presence of membranes with internal structures. In this case we study
mechanisms of localization and normalization for its zero modes and for their Kaluza-Klein modes.

Firstly we introduce in the action of the deformed membrane the
Kalb-Ramond field in the following way
\begin{equation}
\int d^{5}x \sqrt{-G}[2M^{3}R-\frac{1}{2}(\partial\phi_p)^{2}-V_p(%
\phi)-H_{MNL}H^{MNL}],
\end{equation}
where $H_{MNL}=\partial_{[M}B_{NL]}$ is the field strength  for the
Kalb-Ramond field. We will make the gauge choice $B_{\alpha 5}=0$ in
such that its non null components are only those living in the
membrane. We have found the equations of motion for $B_{MN}$ and
made explicit the part dependent on the extra dimension:
\begin{equation}
e^{-2A_p}\partial_{\mu}H^{\mu\gamma\theta}-\partial_{y}H^{y\gamma\theta}=0.
\end{equation}
We make now a separation of variables in order to work the part of
the extra dimension,
\begin{equation}
B^{\mu\nu}(x^{\alpha},y)=b^{\mu\nu}(x^\alpha)U(y)=b^{\mu\nu}(0)e^{ip_\alpha
x^\alpha}U(y),
\end{equation}
where $p^2=-m^2$. We write $H^{MNL}$ as $h^{\mu\nu\lambda}U(y)$. The
equation of motion becomes:
\begin{equation}
\partial_{\mu} h^{\mu\nu\lambda}U(y)-e^{2A_p}\frac{d^2U(y)}{dy^2}b^{\nu\lambda}e^{ip_\alpha x^\alpha}=0.
\end{equation}
The function U(y) carry all information about the extra dimension
and obeys the following equation:
\begin{equation}\label{motion}
\frac{d^2U(y)}{dy^2}=-m^2e^{-2A_p(y)}U(y)
\end{equation}
When $m^2=0$ we have the solutions $U(y)=cy + d$ and $U(y)=c$ with
$c$ and $d$ constants. With these at hand we start to make
computations in order to find localized zero modes of the
Kalb-Ramond field in the deformed membrane. We take the effective
action for the tensor field where we decomposed the part dependent on
the extra dimension,
\begin{eqnarray}
S\sim\int dy U(y)^2
e^{-2A_p(y)} \int d^4 x(h_{\mu\nu\lambda}h^{\mu\nu\lambda}).
\end{eqnarray}\label{effec_action}
Given the solutions for $A_p$ and for $U(y)$ obtained above, we
clearly observe that due to the minus sign in the warp factor, the
function $U(y)^2 e^{-2A_p(y)}$ goes to infinity for the two
solutions of $U(y)$. In this way, the effective action for the zero
mode of the Kalb-Ramond field is not finite after integrating the
extra dimension.

\section{Dilatonic deformed brane}

In the last section, we have not found signals of existence of zero
modes or massive modes trapped to the deformed membrane. The
coupling between the membrane (described by a two-bounce solution)
and the tensor gauge field is strictly due to the space-time metric.
Then, if we want to find localized modes we must modify the
structure of our membrane.

In this point, we would like following the procedure of Refs.\cite{kehagias,kazuo}, where the gauge field localization is
produced by including a new scalar field in the model: the dilaton.
By adding this field in the Einstein equations, we obtain a new
metric behavior and new information about the dynamics of the
membrane. The question here is to understand the behavior of the
Kalb-Ramond field in this new background.

The first step in this analysis is to study the Einstein equations
in this background. We get the action for the membrane, now with two
scalar fields \cite{kehagias},
\begin{equation}
S=\int d^{5}x
\sqrt{-G}[2M^{3}R-\frac{1}{2}(\partial\phi)^{2}-\frac{1}{2}%
(\partial\pi)^{2}-V_p(\phi,\pi)]
\end{equation}
where we denote by $\phi$ the scalar field responsible for the
membrane. The field $\pi$ represents the dilaton. It is assumed a
new ansatz for the spacetime metric:
\begin{equation}
ds^{2}=e^{2A(y)}\eta_{\mu\nu}dx^{\mu}dx^{\nu}+e^{2B(y)}dy^{2}.
\end{equation}
The equations of motion are given by
\begin{equation}
\frac{1}{2}(\phi^{\prime})^{2}+\frac{1}{2}(\pi^{\prime})^{2}-e^{2B(y)}V(%
\phi,\pi)=24M^{3}(A^{\prime})^{2},
\end{equation}
\begin{eqnarray}
\frac{1}{2}(\phi^{\prime})^{2}+\frac{1}{2}(\pi^{\prime})^{2}+e^{2B(y)}V(%
\phi,\pi)=\\\nonumber -12M^{3}A^{\prime\prime}-24M^{3}(A^{\prime})^{2}+12M^{3}A^{\prime}B^{%
\prime},
\end{eqnarray}
\begin{equation}
\phi^{\prime\prime}+(4A^{\prime}-B^{\prime})\phi^{\prime}=\partial_{\phi}V,
\end{equation}
and
\begin{equation}
\pi^{\prime\prime}+(4A^{\prime}-B^{\prime})\pi^{\prime}=\partial_{\pi}V.
\end{equation}
To obtain the first order equations, we choose the following
superpotential $W_p(\phi)$ \cite{kehagias}:
\begin{equation}
V_p=e^{\frac{\pi}{\sqrt{12M^{3}}}}\{\frac{1}{2}\left(\frac{\partial
W_p}{\partial\phi}%
\right)^{2}-\frac{5M^{2}}{2}W_p(\phi)^{2}\}.
\end{equation}
The two kink solutions of the general form (\ref{twokink}) are used in Eq.(\ref{sup}) and we obtain:
\begin{equation}\label{sing}
\pi=-\sqrt{3M^{3}}A_p,\;\;
B=\frac{A_p}{4}=-\frac{\pi}{4\sqrt{3M^{3}}},\;\;
A_p^{\prime}=-\frac{W_p}{3}.
\end{equation}

Contrary to the AdS space-time with negative constant curvature provided by the deformed brane scenario, as we can see in Eq.(\ref{sing}), the solution for
the dilaton makes the space-time singular. The Ricci scalar for this
new geometry is now given by
\begin{equation}
R=-[8A_p''+18(A_p')^{2}]e^{\frac{\pi}{2\sqrt{3M^{3}}}}.
\end{equation}

\begin{figure}
\onefigure[scale=0.6]{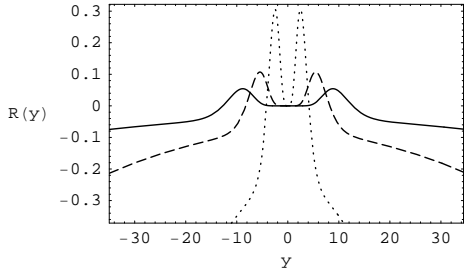}
\caption{Plots of the curvature invariant
$R(y)$ with $p=3,5,7$ (dotted, dashed and solid line 	
respectively).}
\label{fig.3}
\end{figure}

From the Fig.(\ref{fig.3}), we note that there is a region near the membrane
where the Ricci scalar is null. When we take different values for
$p$, the width of this region increases due to the deformation
introduced. Due to the dilaton coupling, the
curvature scalar decreases indefinitely in regions far from the
membrane. However this kind of singularity is very common in D-brane solutions where the dilaton solution is divergent \cite{kehagias}. The work \cite{gremm2} presents some comments about how these singularities can be made harmless.

%Unfortunately, as we can see in Eq.(\ref{sing}), the solution for
%the dilaton makes the space-time singular. The Ricci scalar for this
%new geometry is now given by
%\begin{equation}
%R=-[8A_p''+18(A_p')^{2}]e^{\frac{\pi}{2\sqrt{3M^{3}}}}
%\end{equation}

%\begin{figure}{\centerline
%\epsfig{figure=curvpdil.eps,width=8cm,height=5cm} }
%\caption{Plots of the solution of the curvature invariant
%$R(y)$ with $p=3$ (doted line), $p=5$ (dashed line) and $p=7$ (solid
%line).} \label{fig.3}}
%\end{figure}

\section{Kalb-Ramond field on dilatonic deformed brane}

Now we try again to localize the tensor gauge field, but in the new
background described in the section above. The issue here is to verify if
the dilaton coupling will be enough to localize the Kalb-Ramond
field in the deformed membrane. The dilaton coupling is introduced
in the model in the following way \cite{dilaton1,dilaton2}:
\begin{equation}
S\sim\int d^{5}x(e^{-\lambda\pi}H_{MNL}H^{MNL}).
\end{equation}
Therefore, we must analyze the equations of motion of the tensor
gauge field in the dilaton background. The new equation of motion
is:
\begin{equation}
\partial_{M}(\sqrt{-g}g^{MP}g^{NQ}g^{LR}e^{-\lambda\pi}H_{PQR})=0.
\end{equation}
With the gauge choice $B_{\alpha 5}=\partial\mu B^{\mu\nu}=0$ and
with the separation of variables
$B^{\mu\nu}(x^{\alpha},y)=b^{\mu\nu}(x^{\alpha})U(y)=b^{%
\mu\nu}(0)e^{ip_{\alpha}x^{\alpha}}U(y)$ where $p^{2}=-m^{2}$, it is
obtained a differential equation which give us information about the
extra dimension, namely
\begin{eqnarray}\label{zero}
\frac{d^{2}U(y)}{dy^{2}}-(\lambda\pi^{\prime}(y)+B^{\prime}(y))\frac{dU(y)}{dy}=\\\nonumber-m^{2}e^{2(B(y)-A(y))}U(y)
%\label{U1}.
\end{eqnarray}
For the zero mode, $m=0$, a particular solution of the equation
above is simply $U(y)\equiv cte$. This is enough for the following
discussion. The effective action for the zero mode in $D=5$ is
\begin{eqnarray}
S\sim\int d^{5}x(e^{-\lambda\pi}H_{MNL}H^{MNL})=\\\nonumber \int dy
U(y)^{2}e^{(-2A(y))+B(y)-\lambda\pi(y)}\int d^{4}x(h_{\mu\nu\alpha}h^{\mu\nu\alpha}).
\end{eqnarray}
% WiLAMI%%%%%%%%%%%%%%%%%%%%%%%%%%%%%%%%%%%%%%%%%%%%%%%%%%%%%%%%%%%%%%%%%%%%%%%%%%%%%%%%%%%%%%%%%%%%%%%%%%%%
Given the solution $U(y)$ constant and regarding the solutions for
$A_p(y)$, $B(y)$ e $\pi(y)$, it is possible to show clearly that the
integral in the $y$ variable above is finite if $\lambda >
\frac{7}{4\sqrt{3M^3}}$, and  for $p$ finite. As a consequence, for
a specific value of the coupling constant $\lambda$ it is possible
to obtain a localized zero mode of the Kalb-Ramond field.
%%%%%%fim-Wilami|%%%%%%%%%%%%%%%%%%%%%%%%%%%%%%%%%%%%%%%%%%%%%%%%%%%%%%%%%%%%%%%%%%%%%%%%%%%%%%%%%%%%%

\section{Massive modes}

We should now consider a discussion about massive modes in this
background. For this, we must analyze the Eq. (\ref{zero}) for
$m\neq0$ trying to write it in a Schroedinger-like equation through
the following change
\begin{equation}\label{trans2}
dz=dy e^{-\frac{3}{4}A_p },\;\;\;  U=e^{\left(\frac{\alpha}{2}+\frac{3}{8}\right)A}\overline{U},\;\;\;
\alpha=\frac{1}{4}-\sqrt{3M^3}\lambda.
\end{equation}
After all the necessary calculations we arrive at the equation we
want to analyze, namely
\begin{equation}\label{schro}
\left\{-\frac{d^2}{dz^2}+\overline{V}(z)\right\}\overline{U}=m^2\overline{U},
\end{equation}
where the potential $\overline{V}_p(z)$  assumes the form,
\begin{equation}\label{vp}
\overline{V}_p(z)=\left[\beta^2(\dot{A}_p)^2-\beta\ddot{A}_p\right],\;\;\;\;
\beta=\frac{\alpha}{2}+\frac{3}{8}.
\end{equation}

%\begin{figure}{\centerline
%{\epsfig{figure=volkdil.eps,width=7cm,height=6cm} }
%\caption{Plots of the potential $\overline{V}_p(z)$  with $p=3$ (dashed line), $p=5$ (doted %line) and $p=7$ (solid line). We have put $\sqrt{3M^3}\lambda=2$. } \label{}}
%\end{figure}

\begin{figure}
\onefigure[scale=0.50]{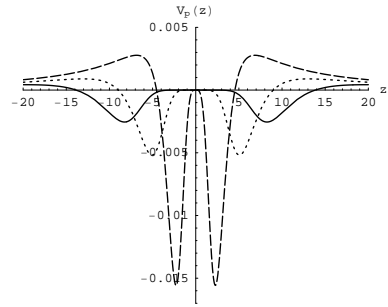}
\caption{Plots of the potential $\overline{V}_p(z)$  with $p=3$ (dashed line), $p=5$ (doted line) and $p=7$ (solid line). We have put $\sqrt{3M^3}\lambda=2$.}
\label{fig.30}
\end{figure}

We can see from the Fig.(\ref{fig.30}) that the potential is
affected by the deformation procedure introduced in this work. We
identify the existence of two minima whose distance increases when
we increase the values of $p$. The form of the potential is also directly
changed. Note that we use $\sqrt{3M^3}\lambda > 1$  in order to
obtain a potential like (\ref{vp}), i.e., the standard form found
when we write Schroedinger-like equations. This choice is
fundamental in order to find finite results regarding the behavior
of the Kalb-Ramond field.

It is interesting to point out that the Schrodinger-type equation
(\ref{schro}) can be written in the supersymmetric quantum mechanics
scenario as follows,
\begin{equation}\label{susy_qm}
Q^{\dag} \, Q \,
\overline{U}(z)=\left\{\frac{d}{dz}-\beta\dot{A}\right\}\left\{\frac{d}{dz}+\beta\dot{A}\right\}\overline{U}(z)=-m^2\overline{U}(z).
\end{equation}
From the form of the Eq. (\ref{susy_qm}), we exclude the possibility
of normalized negative energy modes to exist. On the other hand, we
exclude also the possibility of the presence of tachyonic modes,
which is a necessary condition to keep the stability of
gravitational background.

\begin{figure}
\onefigure[scale=0.85]{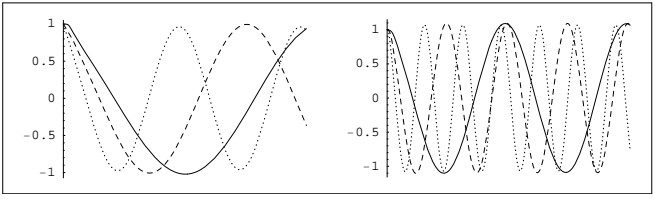}
\caption{ Plots of $\overline{U}_p(z)$ for $p=1$ (doted
line), $p=3$ (dashed line), and $p=5$ (solid line), where we put $m^2 \leq {\overline{V}_p}(z)_{max}$ (left) and
$m^2 > {\overline{V}_p}(z)_{max}$ (right). We use $\sqrt{3M^3}\lambda=2$.}
\label{wavepdil}
\end{figure}

We cannot find analytical solution of the massive modes wave
function in Schrodinger equation. However we will be able to analyze
the solution for $\overline{U}_p$ by numerically solving the
equation (\ref{schro}). We plot in Fig.(\ref{wavepdil}) the wave
function so obtained for two values of $m^2$. As we can observe,
when we make $m^2 > \overline{V}_p(z)_{max}$, we minimize the
contribution due to the deformations over the solution
$\overline{U}_p$. However, regarding $m^2 \leq
{\overline{V}_p}(z)_{max}$, as we increase $p$ we reduce the
frequency of oscillations of the solutions $\overline{U}_p$. We must
remember that the search for finite solutions it was only possible due
to the choice $\sqrt{3M^3}\lambda> 1$.

%\begin{figure}{\centerline{
%\epsfig{figure=wavepdil.eps,width=15.2cm,height=4.2cm}}
%\caption{ Plots of $\overline{U}_p(z)$ for $p=1$ (doted
%line), $p=3$ (dashed line), $p=5$ (solid line) and $p=7$ (thick
%line), where we put $m^2 \leq {\overline{V}_p}(z)_{max}$ (left) and
%$m^2 > {\overline{V}_p}(z)_{max}$ (right)}\label{wavepdil}}
%\end{figure}

%+++++++++++++++++início++++++++++++++++++++

As mentioned in Ref. \cite{adauto}, the behavior of the
wave function suggests us a free motion in the bulk, but no trapping
in the membrane.
%++++++++++++++++++++++++++++++++++++++++++++++++++++++++++++++
\begin{figure}
\onefigure[scale=0.85]{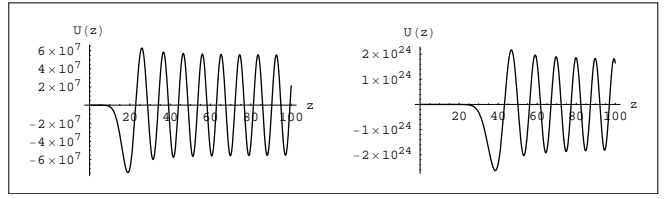}
\caption{ Plots of $\overline{U}_p(z)$ for $p=1$, where we put $m=0,5$, $\sqrt{3M^3}\lambda=20$ (left) and $\sqrt{3M^3}\lambda=40$ (right).}
\label{dil}
\end{figure}
An interesting point to be investigated is how the intensity of the
dilaton constant coupling may modify the patterns of solutions
observed in Fig.(\ref{wavepdil}). For this, we solve again the
equation (\ref{schro}), but this time, changing the values of the
constant coupling $\lambda$. We plot in Fig.(\ref{dil}) the
function $\overline{U}_p(z)$, this time for $\sqrt{3M^3}\lambda=20$
on the left and $\sqrt{3M^3}\lambda=40$ on the right. We note the
suppression of the mode oscillations in regions near the membrane
due to the increasing of $\lambda$. On the other hand, in regions far
away from $z=0$ the amplitude oscillations grows.

%\begin{figure}{\centerline{ \epsfig{figure=dil.eps,width=14.2cm,height=4cm}},\caption{Plots of %$\overline{U}_p(z)$ for $p=1$, where we put $m=0,5$, $\sqrt{3M^3}\lambda=20$ (left) and %$\sqrt{3M^3}\lambda=40$ (right).} \label{dil}}
%\end{figure}

\section{Resonances}
\begin{figure}
\onefigure[scale=0.6]{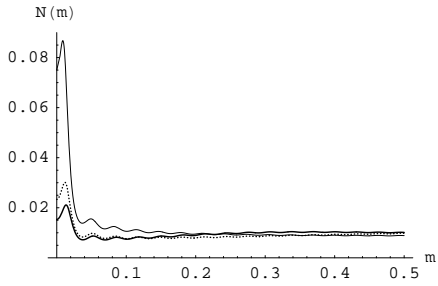}
\caption{ Plots of $N_p(m)$ for $p=1$ (thin line), $p=3$ (points) and $p=5$ (thick line). We use $\sqrt{3M^3}\lambda=2$.}
\label{resop}
\end{figure}

The study of the resonances is important since it can gives relevant information about the massive modes \cite{ca,liu1,liu2,bazeia1,cvetic,liu3}.
In order to better understand the coupling between massive modes and
matter on the membrane we should know, starting from Eq.(\ref{susy_qm}), the amplitude of the plane wave function
$\overline{U}_p(z)$ normalized at $z=0$. As is pointed out in Refs. \cite{gremm} and \cite{csaba}, for highly
massive modes in relation to $\overline{V}(z)_{max}$, the potential
represents only a little perturbation. Nevertheless, it is possible
that modes of the function $\overline{U}(z)$ for which
$m^2\leq\overline{V}(z)_{max}$ can resonate with the potential. The quantity
$|\zeta\overline{U}_m(0)|^2$, being $\zeta$ a normalization
constant, should give us the probability of finding a mode of mass
$m$ at $z=0$ and is given by
\begin{equation}
N_p(m)=\frac{|\overline{U}_m(0)|^2}{\int_{-100}^{100}|\overline{U}_m(z)|^2 dz}
\end{equation}
In the Fig.(\ref{resop}) we plot
$N_p(m)=|\zeta\overline{U}_p(0)|^2$ where we can identify, for
$p=1$, a resonant peak near $m=0$, precisely for $m=9\times
10^{-3}$. We may interpret that, in this case, the probability of
finding light modes or massless modes coupled to the membrane is
bigger than for heavier modes. This characteristics disappears when
we change the values of $p$. As we can observe in Fig.(\ref{resop}), the resonant structure tends to disappear in
accordance to the results of localization of the zero mode.
%\begin{figure}{\epsfig{figure=resop.eps,width=8cm}
%\caption{Plots of $M_p(m)$ for $p=1$ (thin line), $p=3$ (points) and $p=5$ (thick line). %}\label{resop}}
%\end{figure}

\begin{figure}
\onefigure[scale=0.6]{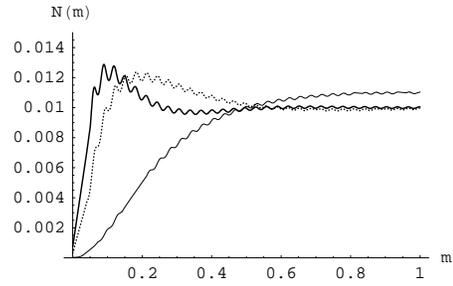}
\caption{ Plots of $N_p(m)$ for $p=1$ (thin line), $p=3$ (points) and $p=5$ (thick line). We use $\sqrt{3M^3}\lambda=0$.}
\label{reso2p}
\end{figure}

We can still test the consistency of the above results regarding again the
model without the dilaton coupling. In this
case, we do not find signals of localization of the Kalb-Ramond
field. In this way, supressing the dilaton couling, we can extract the function $N_p(m)$ from
equation (\ref{schro}) by the same steps discussed before and plot the
results in Fig.(\ref{reso2p}). As we expect, the resonant
structure disappears and the couplings of the zero modes is highly
suppressed.

%\begin{figure}{\epsfig{figure=reso2p.eps,width=8cm} \caption{Plots of $M_p(m)$ for $p=1$ (thin %line), $p=3$ (points) and $p=5$ (thick line).}\label{reso2p}}
%\end{figure}

\section{Conclusions}

In this article we analyze under several aspects the localization
properties of the Kalb-Ramond tensor gauge
field in a very specific type of membrane: the deformed membrane.

The analysis of the Kalb-Ramond field is jeopardized since the effective action is not normalizable: we have not zero modes for the Kalb-Ramond field. The resulting equation of motion for the massive modes is not found and it can not be written in a form of Schrodinger-like equation. This fact do not allows us to interpret quantum mechanically the problem. What we do to circumvent this result is to add one more field in the model, the dilaton, and this changes a little the gravitational background. After this modification, we can, under some conditions, find a localized tensorial zero mode. Related to the spectra of massive states, we see that the effective potential in the Schrodinger-like equation is affected by the deformations made in the membranes. The numerical analysis of that equation for massive states reveals that there are plane waves describing the free propagation of particles in the bulk. The dilaton coupling change the amplitude of oscillations of the modes away from the membrane. Indeed, studying the coupling of the matter massive states with the membrane we have found a resonance, which again disappears with the deformations. The resonance structures show us that only light modes of the KK spectrum present not suppressed coupling with the membrane. Finally, we showed the consistency of the results obtained with those from the model without the dilaton.

\acknowledgments
The authors would like to thank Funda\c{c}\~{a}o Cearense de apoio ao Desenvolvimento
Cient\'{\i}fico e Tecnol\'{o}gico (FUNCAP) and Conselho Nacional de Desenvolvimento
Cient\'{\i}fico e Tecnol\'{o}gico (CNPq) for financial support.

\end{document}